\documentclass[structabstract]{aa}

\usepackage{graphicx}
\usepackage{txfonts}
\usepackage{natbib}
\usepackage{hyperref}

\begin{document}
   \title{Reflection nebulae in the Galactic Center: the case for soft X-ray imaging polarimetry}
   \titlerunning{Soft X-ray polarization maps of the GC}

   \author{F. Marin\inst{1}\thanks{\email{frederic.marin@asu.cas.cz}}
	    \and F.~Muleri\inst{2} 
	    \and P.~Soffitta\inst{2} 
	    \and V.~Karas\inst{1}
	    \and D.~Kunneriath\inst{1}}
	     
   \institute{Astronomical Institute of the Academy of Sciences, 
     Bo{\v c}n\'{\i} II 1401, CZ-14100 Prague, Czech Republic
     \and 
     INAF/IAPS, Via del Fosso del Cavaliere 100, I-00133 Roma, Italy}

   \date{Received November 14th 2014 / Accepted February 17th 2015}


  \abstract
  {Despite numerous attempts with spectroscopic and timing analyses, the origin of irradiation and fluorescence of 
    the 6.4~keV bright giant molecular clouds surrounding Sgr~A$^*$, the central supermassive black hole of our 
    Galaxy, remains enigmatic.} 
  {Testing the theory of a past active period of Sgr~A$^*$ requires to open a new observational window: 
    X-ray polarimetry. In this paper, we aim to show how modern imaging polarimeters could revolutionize our 
    understanding of the Galactic Center.}
  {Through Monte Carlo modeling, we produce a 4 -- 8~keV polarization map of the Galactic Center, focusing on the 
    polarimetric signature produced by Sgr~B1, Sgr~B2, G0.11-0.11, Bridge~E, Bridge~D, Bridge~B2, MC2, MC1,
    Sgr~C3, Sgr~C2, and Sgr~C1. We estimate the resulting polarization arising from those scattering targets, 
    include polarized flux dilution by the diffuse plasma emission detected toward the GC, and simulate the 
    polarization map that modern polarimetric detectors would obtain assuming the performances of a mission prototype.}
  {The eleven reflection nebulae investigated in this paper present a variety of polarization signatures, ranging 
    from nearly unpolarized to highly polarized ($\sim$ 77\%) fluxes. Their polarization position angle is found 
    to be normal to the scattering plane, as expected from previous studies. A major improvement in our simulation
    is the addition of a diffuse, unpolarized plasma emission that strongly impacts soft X-ray polarized fluxes.
    The dilution factor is in the range 50\% -- 70\%, making the observation of the Bridge structure unlikely even in 
    the context of modern polarimetry. The best targets are the Sgr~B and Sgr~C complexes, and the G0.11-0.11 cloud,
    arranged in the order of decreasing detectability.}
  {An exploratory observation of a few hundred kilo-seconds of the Sgr~B complex would allow a significant detection 
    of the polarization and be sufficient to derive hints on the primary source of radiation. A more ambitious program 
    (few Ms) of mapping the giant molecular clouds could then be carried out to probe with great precision the turbulent 
    history of Sgr~A$^*$, and place important constraints on the composition and three-dimensional position of the surrounding gas.}


\maketitle

\section{Introduction}
\label{Intro}

Using the {\it Herschel} satellite, \citet{Molinari2011} recently discovered a massive ($\sim$ 3$\times$10$^{7}$ 
M$_{\odot}$), continuous chain of irregular, cold dusty clumps in the vicinity of Sgr~A$^*$, the central supermassive 
black hole (SMBH) of the Milky Way. The thermal, far-infrared images obtained reveal a $\infty$-shaped, twisted	 
ring that is reminiscent of the persistent dusty tori surrounding the central regions of active galactic nuclei 
(AGN). In addition, the geometrical size of the circumnuclear gas structure, its column density in excess 
of 10$^{24}$~cm$^{-2}$, and its orbital speed ($\sim$100 km.s$^{-1}$, \citealt{Molinari2011}), are compatible 
with AGN tori \citep{Shi2006}. Yet, the current quiescent X-ray luminosity of Sgr~A$^*$ ($L_X~\sim$~2~$\times$~10$^{33}$~ergs.s$^{-1}$,
\citealt{Baganoff2001}) is orders of magnitude lower than what is observed in Seyfert-1 AGN ($L_X~>$~10$^{40}$~ergs.s$^{-1}$), 
where high accretion rates (typically 0.01 to 0.2 M$_{\odot}$.y$^{-1}$, \citealt{Meyer2011}) provide efficient radiating 
engines. Therefore, the question of a more turbulent history, i.e. an active phase, of Sgr~A$^*$ becomes of prime interest.

It has been suggested that the central SMBH underwent at least two high-luminosity periods, bright enough to 
illuminate its environment \citep{Inui2009,Ponti2010}. Traces of this potential activity can be found from
the epoch of \textit{Granat}, when \citet{Sunyaev1993} provided broadband 15' resolution images of the Galactic Center 
(GC). In their observations, the GC is characterized by a spherical shape in the 2.5 -- 5~keV X-ray band and by an extended 
(i.e. elongated along the Galactic plane) morphology in the 8.5 -- 19~keV energy range. To explain such a difference in 
the spatial structure of the GC emission, \citet{Sunyaev1993} suggested that part of the diffuse emission of the 
molecular gas clouds, associated with very steep spectra and strong iron fluorescent emission lines \citep{Koyama1996}, 
may be due to Compton scattering of photons originating from a nearby compact source. Additional detections of hard X-ray 
spectral slopes and Fe~K$\alpha$ emission lines from a variety of neighboring GC gas clouds \citep{Murakami2001b,Ponti2010,Capelli2012} 
strengthened the classification of a tenth of giant molecular clouds as reflection nebulae, echoing past Sgr~A$^*$ outbursts.

The spatial position of the reflectors becomes crucial in the process of determining the goodness of the flaring theory 
(with estimated $L_X >$~10$^{39}$~erg.s$^{-1}$). \citet{Churazov2002} proved that a polarimetric mission, inherently 
sensitive to the morphology and the location of reprocessing targets, is the most adequate solution to investigate
the re-emitted flux of the scattering molecular clouds. In their model, the Sgr~B2 cloud is expected to produce a high
polarization degree associated with a direction of polarization normal to the scattering plane. A more elaborate investigation 
has been undertaken in \citet{Marin2014}, where we produced 8 -- 35~keV polarization maps of the GC. We avoided the soft X-ray 
energies, since past X-ray observations \citep{Koyama1986,Koyama1989,Sidoli1999} have revealed the presence of a diffuse plasma 
emission angularly superimposed to the X-ray emission of the molecular clouds. This diffuse emission can be well-explained with 
a two-temperature plasma with T$_{\rm 1}\le$ 1~keV and T$_{\mathrm{2}}$ = 5 -- 7~keV \citep{Koyama2007}. It is probably 
due to a multitude of faint sources (accreting white dwarfs and coronally active stars, \citealt{Revnivtsev2009}). This 
X-ray component should be basically unpolarized \citep{Mewe1999}, ultimately diluting the polarization signal at energies $E$ 
below 7~keV. Avoiding those energies in \citet{Marin2014}, we conservatively modeled the Sgr~B2 cloud following the prescription 
by \citet{Churazov2002}. We also implemented a simple structure for the Sgr~C complex, as well as the dusty, twisted ring discovered 
by \citet{Molinari2011}, and a reservoir of gas surrounding the inner 5~pc around Sgr~A$^*$ (not to be mistaken for an accretion disk). 
It was found that only the two reflection nebulae can be detected at high energies, but it is unknown if similar results hold 
at $E~\le$~7~keV.

It is the scope of this paper to extend the investigation of \citet{Churazov2002} and \citet{Marin2014} to the soft X-ray band,
by increasing the number of reflection nebulae in the model, and estimating the plasma and the reflected contributions for the molecular 
clouds in order to produce a realistic, 4 -- 8~keV polarization map of the GC. In a crowded field such as the GC, the presence of an 
{\em imaging} detector becomes necessary to resolve the faint gas clouds and probe the scattering pattern of radiation.
To precisely localize the reflection nebulae, characterize their composition and reveal the past activity of Sgr~A$^*$, we present in 
Sect.~\ref{Main:Model} the Monte Carlo simulations we performed to obtain a synthetic polarimetric image of the GC. We estimate the 
polarized flux dilution by the diffuse plasma emission detected toward Sgr~A$^*$ and compute the diluted polarization signal that a 
modern imaging polarimeter could detect from space in Sect.~\ref{Main:IXPE}. We discuss our results and conclude our paper in 
Sect.~\ref{Conclusion}.

\section{A polarimetric, soft X-ray, view of the Galactic Center}
\label{Main}

\subsection{Modeling the polarization from reflection nebulae}
\label{Main:Model}

\begin{figure}
   \centering
      \includegraphics[width=10.5cm]{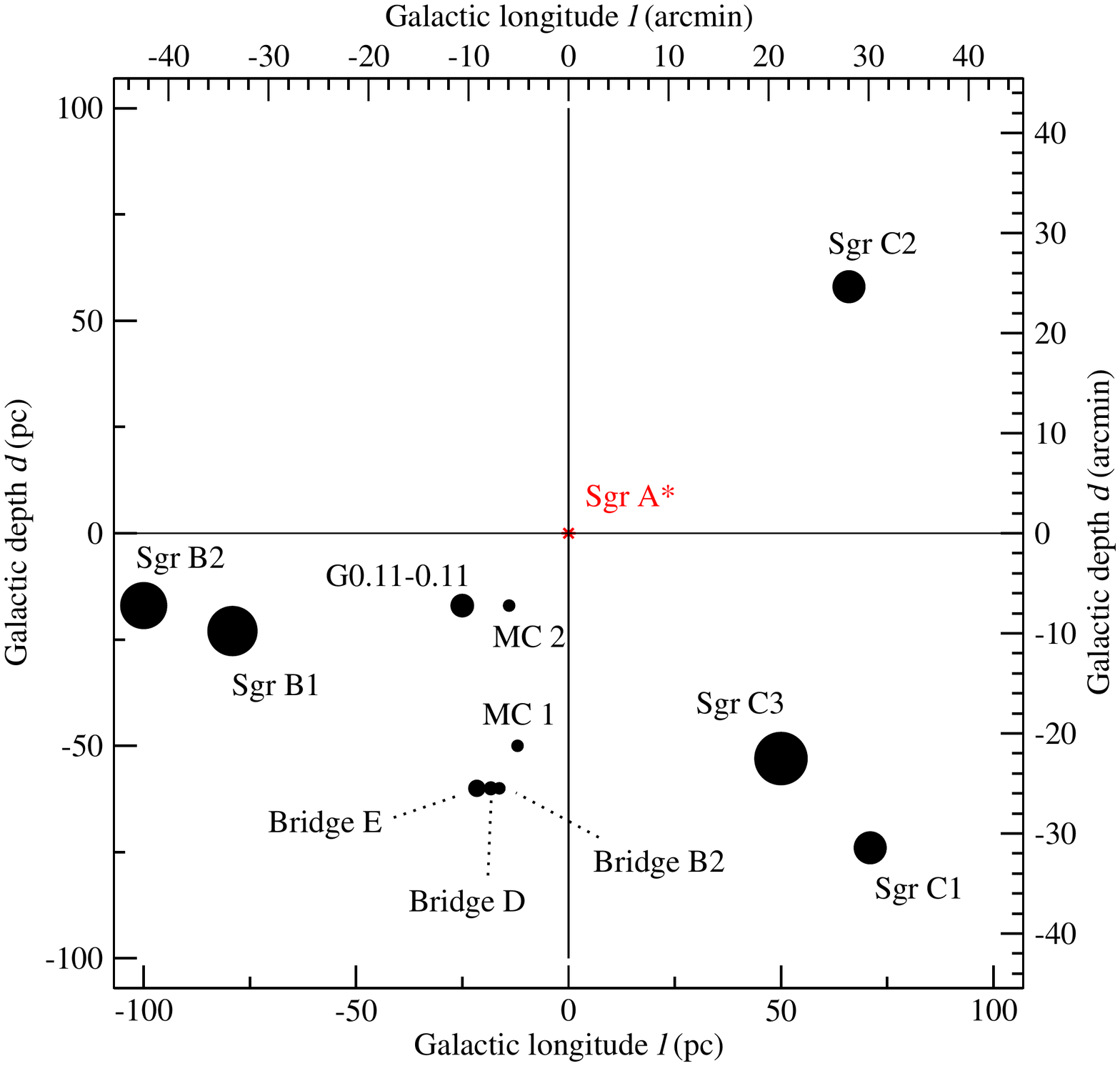}
      \includegraphics[width=10.5cm]{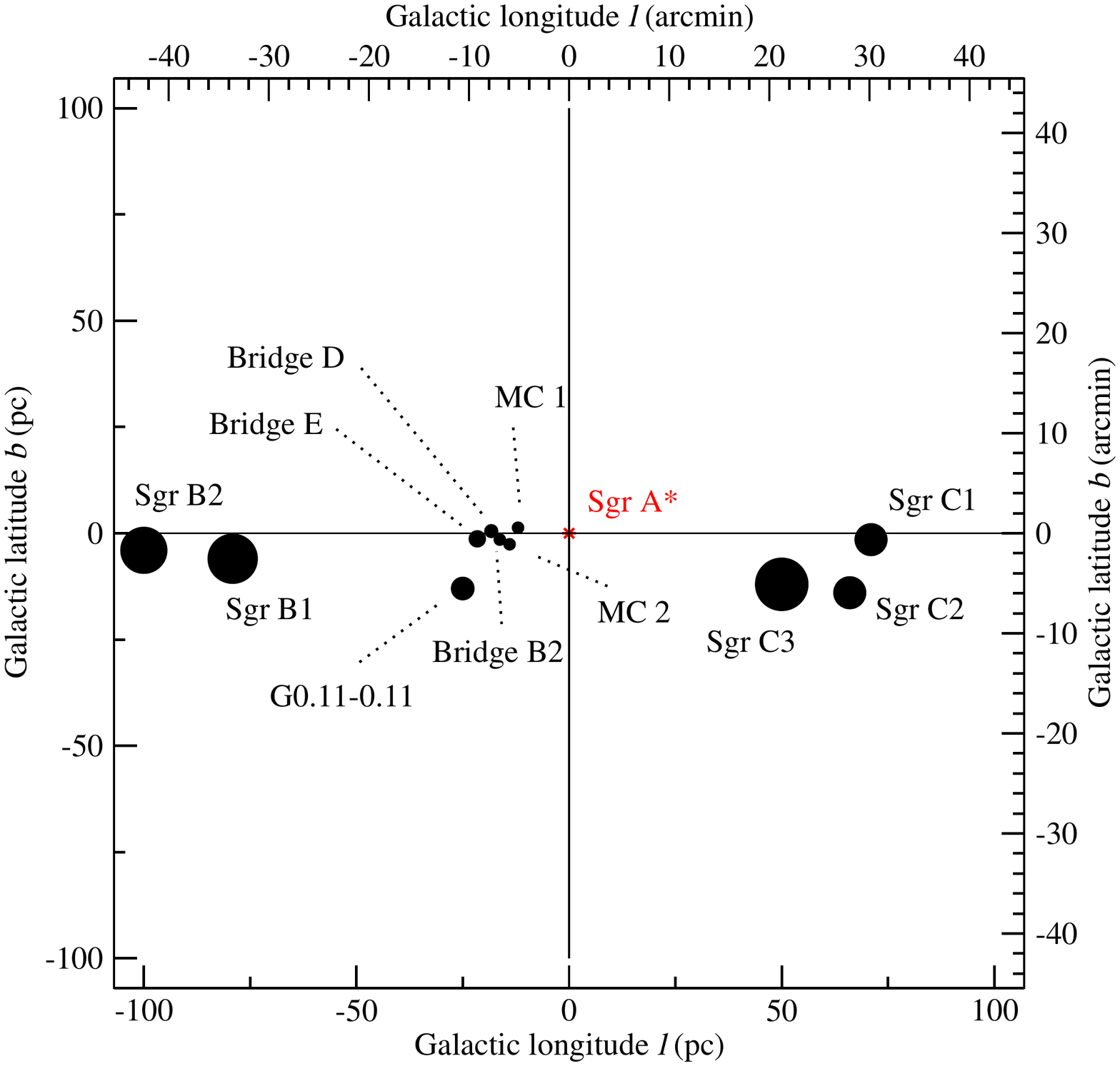}
      \caption{Sketch of the GC model as seen from two directions. Top: view from the direction of 
      the Galactic pole; the Earth is situated toward negative Galactic depths. Bottom: view 
      from Earth; each cloud is projected on the plane of the sky (the Galactic plane).}
     \label{Fig:Sketch}
\end{figure}

\begin{table*}
  \begin{center}
    \begin{tabular}{p{2cm} p{1cm} p{1cm} p{1.7cm} p{1cm} p{1cm} p{3cm} p{1.5cm} p{1cm}}
    \hline
    \hline
	Molecular cloud & Cloud radius (pc) & Projected distance$^{\rm a}$ (pc) & Line of sight distance$^{\rm b}$ (pc) & 
	Offset$^{\rm c}$ (pc) & Velocity$^{\rm d}$ (km.s$^{-1}$) & Hydrogen column density ($\times$ 10$^{22}$ cm$^{-2}$) & Electron optical depth & References \\
    \hline
	Sgr~B2 & 5 & -100 & -17 & -4.0 & 60 & 80 & 0.5 & E,I\\
	Sgr~B1 & 6 & -79.1 & -23 & -6 & -45 & 12.3 & 0.3 & A,D,G\\
	G0.11-0.11 & 3.7 & -25 & -17 & -13 & 25 & 2 & 0.03 & E,F\\
	Bridge~E & 2.0 & -21.6 & -60 & -1.3 & 55 & 9.6 & 0.07 & B,E,F\\
	Bridge~D & 1.6 & -18.3 & -60 & 0.5 & 55 & 13.2 & 0.09 & B,E,F\\
	Bridge~B2 & 1.8 & -16.3 & -60 & -1.5 & 55 & 12.3 & 0.08 & B,E,F\\
	MC2 & 1.8 & -14 & $<$-17 & -2.6 & -10 & $<$2 & 0.36 & C,E\\
	MC1 & 1.8 & -12 & -50 & 1.3 & -15 & 4 & 0.32 & E\\
	Sgr~C3 & 6 & 50 & -53 & -12 & 60 & 8.7 & $<$1 & H,E\\
	Sgr~C2 & 4.7 & 66 & 58 & -14 & 60 & 11.4 & $<$1 & H,E\\
	Sgr~C1 & 4.7 & 71 & -74 & -1.5 & 60 & 6.5 & $<$1 & H,E\\
    \hline
    \end{tabular}
    \caption{Parameterization of the reflection nebulae, modeled with uniform-density, 
              spherical clouds filled with cold, solar abundance matter. $^{\rm a}$Positive = East of the Galactic center; 
	      $^{\rm b}$Positive = behind the Galactic plane (farther to us than Sgr~A$^*$); 
	      $^{\rm c}$Positive = above the equatorial plane.
	      $^{\rm d}$Positive = away from Earth.
	      References -- A: \citet{An2013}; B: \citet{Capelli2012}; C: \citet{Clavel2013}; 
	      D: \citet{Downes1980}; E: \citet{Ponti2010}; F: \citet{Ponti2014}; 
	      G: \citet{Ryu2009}; H: \citet{Ryu2013} and I: \citet{Sunyaev1993}.}
  \end{center}
  \label{Tab:Model} 
\end{table*}

We model the past activity of Sgr~A$^*$ as a point-like accreting source at the location of the SMBH, emitting an unpolarized 
spectrum with a spectral energy distribution $F_{\rm *}~\propto~\nu^{-\alpha}$ ($\alpha = 1.0$, \citealt{Porquet2003,Porquet2008,Nowak2012}). 
The resulting 4 -- 8~keV emission is isotropic and photons journey through the model until absorption/reemission/scattering onto 
the giant molecular gas clouds. Polarization of the observed signal then arises from Compton scattering of the reprocessed light, 
where the scattering angle determines the polarization degree and the polarization angle of the intercepted signal that can be 
recorded at the detector. The reflection nebulae are modeled with uniform-density, spherical clumps filled with neutral solar 
abundance matter and located according to the most recent constraints from infrared-to-X-rays observations (see Tab.~1). 
A sketch of the model is presented in Fig.~\ref{Fig:Sketch}, showing the location of the reflection nebulae from a polar view 
(top figure) and on the plane of the sky (i.e. the Galactic plane, bottom figure). The axes are labeled in parsecs and arcminutes.

\begin{figure*}
  \centering
  \includegraphics[trim = 15mm 65mm 0mm 50mm, clip, width=19.5cm]{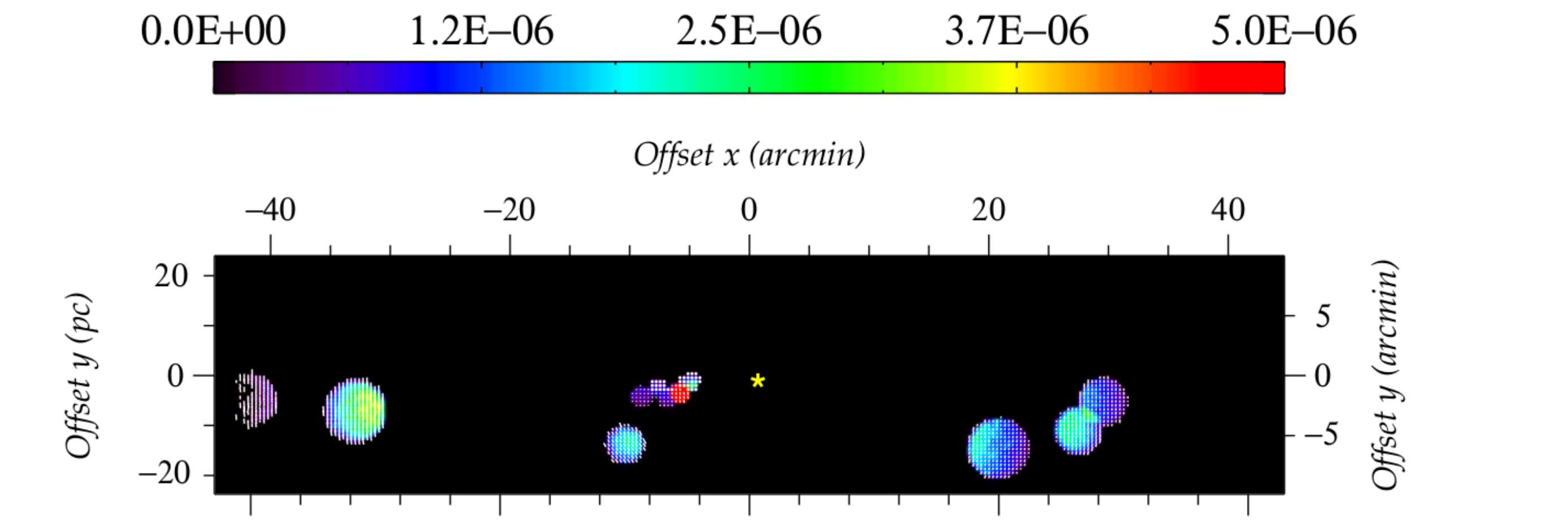} 
  \includegraphics[trim = 15mm 65mm 0mm 90mm, clip, width=19.5cm]{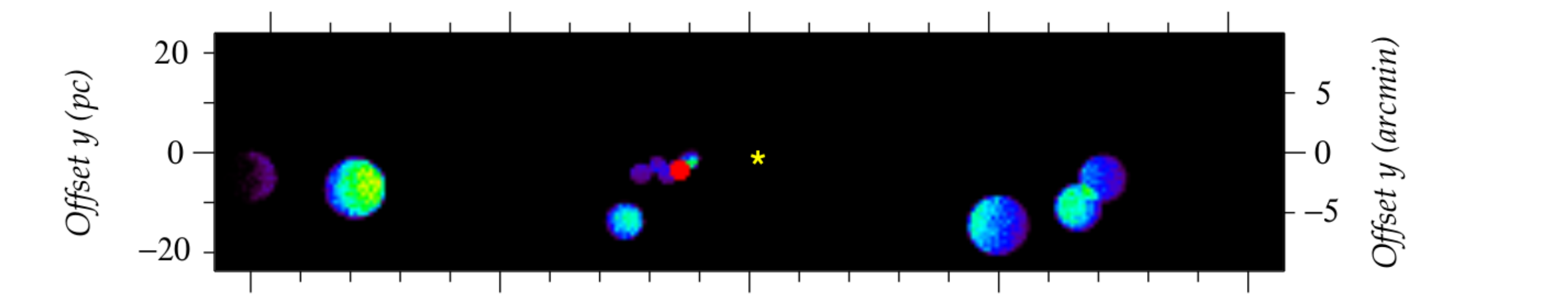} 
  \includegraphics[trim = 15mm 50mm 0mm 90mm, clip, width=19.5cm]{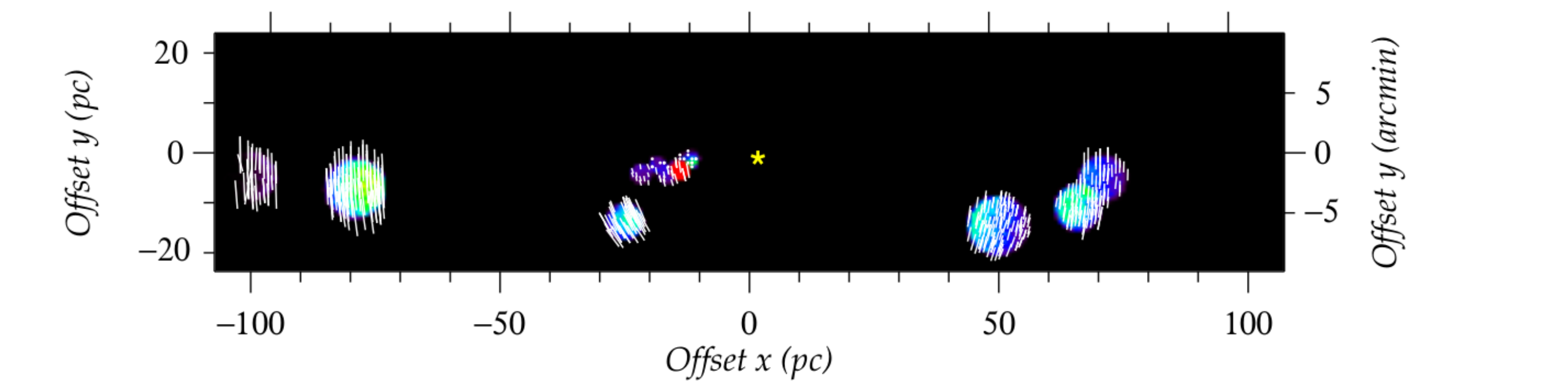} 
  \includegraphics[trim = 15mm 50mm 0mm 45mm, clip, width=19.5cm]{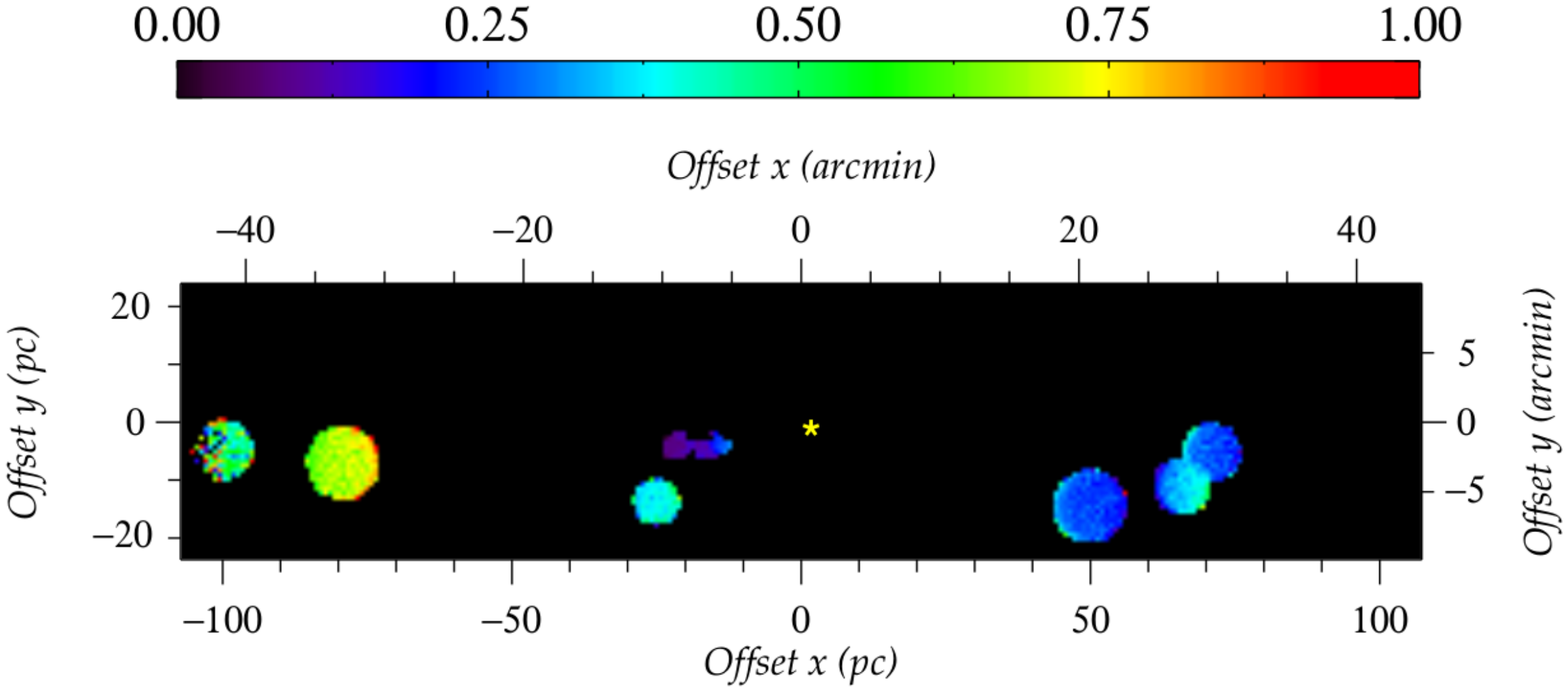} 
  \caption{Simulated model images of the $\sim$~100~pc $\times$ 20~pc region around 
	   Sgr~A$^*$. The top map shows the combination of polarized flux, 
	   $PF/F_{\rm *}$ (color-coded, with the color scale shown on top of the 
	   image in arbitrary units), polarization position angle $\psi$ 
	   (white bars drawn in the center of each spatial bin), and 
	   polarization degree $P$, whose value is proportional to the length 
	   of the bars. The second map is the polarized flux only, the third image
	   the polarization position angle $\psi$ with artificially extended white 
	   vectors for better visibility, and the bottom map represents the 
	   $P$, color-coded, with the color scale shown on top of the image 
	   in fraction of polarization. A yellow star indicates the 
	   position of Sgr~A$^*$.} 
  \label{Fig:Modeling} 
\end{figure*}

Three-dimensional radiative transfer is achieved using {\sc stokes} \citep{Goosmann2007,Marin2012}, a Monte Carlo code that includes 
a coherent treatment of polarization, multiple scattering, and an imaging routine. Computation of the re-emitted spectra includes 
algorithms for inelastic Compton scattering onto bound electrons, photo-absorption and iron line fluorescent. The emission direction, 
the distance that photons travel between reprocessing events, and the scattering angles are computed by Monte Carlo routines based on classical 
intensity distributions. Mueller matrices are used to evaluate the change in polarization after each scattering event. Photo-absorption above the 
atom K-shell and the subsequent emission of K$\alpha$ and/or K$\beta$ line photons is included and weighted against the probability of Auger 
effects. For further details about the code, please refer to the complete description of the polarization properties and transformation of radiation 
during scattering events described in \citet{Goosmann2007}, \citet{Marin2012} and \citet{Marin2015}.

We sample a total of 7$\times$10$^{11}$ photons in a model with a spatial resolution set to 270$\times$270~bins for the longitudinal and latitudinal 
offsets, so that the photon flux is divided into 72900 pixels. Each of these pixels is labeled by its position offset in parsecs and arcminutes, and 
stores the four Stokes parameters of the photons. The spatial resolution is equal to 0.8~pc, which represents 20~arcsecs at the distance of the GC 
(8.5~kpc, \citealt{Ghez2008}). Finally, the model space is divided in 20 polar and 10 azimuthal viewing directions. Note that due to the three-dimensional 
meshes of the coordinate grid, the shape of the scattering regions is slightly deformed in the image projection process.

The resulting polarization maps of the GC, integrated over the whole 4 -- 8~keV band to maximize detection, are presented in Fig.~\ref{Fig:Modeling}. 
The top figure shows a triple combination of 1) the polarized flux ($PF/F_{\rm *}$, i.e. intensity $\times$ polarization degree), color-coded and 
displayed with arbitrary units, 2) the polarization degree $P$, and 3) the polarization position angle $\psi$ identified by white bars drawn in the 
center of each spatial bin. A vertical bar indicates a polarization angle of $\psi$~=~90$^\circ$ and a horizontal bar stands for an angle of 
$\psi$~=~0$^\circ$. The length of the bar is proportional to $P$. The second figure shows the polarized  flux only and the third is a visual 
representation of $\psi$ with artificially extended white vectors for better visibility. The fourth, bottom, map depicts the polarization degree 
with its own color code, ranging from 0 (unpolarized) to 1 (fully polarized).

From East to West, we find that Sgr~B2 presents a high polarization degree (65.0~\%) associated with very small polarized fluxes (a consequence 
of its large hydrogen column density and distance from the irradiating source). The polarized flux map (Fig.~\ref{Fig:Modeling}, top) clearly 
shows a brightness distribution of the flux on the contours of the molecular gas model that is facing  the SMBH, such as observed by \citet{Murakami2001a}. 
Sgr~B1 has the highest polarization degree of the GC, up to 76.9~\%. Its size and location allow a large polarized flux to be 
observed. Similarly to the other big structures, the re-emission pattern from the cloud can be probed in great detail by imaging polarimeters.
Similarly, G0.11-0.11 shows large polarized fluxes due to a reasonably high $P$ (55.8~\%). The Bridge globally displays medium-to-low polarized 
fluxes. The three-dimensional location of the clouds forming the Bridge (Bridge~D, Bridge~E, Bridge~B2, MC1 and MC2) explains their lower polarization 
degrees (from 0.06 to 15~\%) in comparison with the other scattering nebulae (see Fig.~\ref{Fig:Sketch}). One notable exception is the MC2 cloud, 
exhibiting a polarization degree up to 25.8~\% since, being the closest cloud to Sgr~A$^*$, its scattering angle with respect to the source and the 
observer is more favorable. The $\sim$~10~\% polarization of MC2's neighboring, coplanar clouds (Bridge~B2 and E) arises from scattering 
of high-$P$ photons reprocessed on MC2 and then reaching the observer. Finally the Sgr~C complex behaves uniformly despite the dispersion of 
its three clouds with respect to the line-of-sight distance. They exhibit moderate polarized fluxes and polarization degrees of the order of 32~\%. 
All the clouds display a polarization position angle $\psi$ normal to the scattering  plane (i.e. close to 90$^\circ$). We summarize the integrated 
$P$ and $\psi$ in the first two columns of Tab.~2.

Thus, the GC presents a large panel of polarization signatures associated with polarization degrees varying from high\footnote{High degrees
of integrated polarization are reachable despite the presence of an unpolarized iron fluorescence line at 6.4~keV. The amount of dilution 
depends on the strength and equivalent width of the line: for a 1~keV equivalent width (as for Sgr~B2, \citealt{Sunyaev1998}), the 
line flux counts for about 20\% of the total flux in the 4 -- 8~keV band so the dilution of polarization due to this line is small.}
(76.9~\%) to very low values (0.1~\%). The blend of the polarization signals originating from the Sgr~B and Sgr~C complexes, and 
from the Bridge structure, underlines the need for an imaging detector with a sufficient spatial resolution in order to resolve structures 
as small as the Bridge clouds. Additionally, our results are found to be consistent with the pioneering simulation of \citet{Churazov2002} 
and their higher energy counterpart \citep{Marin2014}. However, in the light of our previous (8 -- 35~keV) simulations \citep{Marin2014}, 
it is worth mentioning that our polarization results strongly depend on the real location of the reflection nebulae. As it was shown in the aforementioned
article, the degree of polarization resulting from reprocessing onto the outer layers of the cloud approximately varies as the square of the 
cosine of the scattering angle between the source, the cloud, and the observer's position. With new estimations of the true location of the 
scattering nebulae, first order corrections can be then applied to results from Tab.~2 (see, e.g., discussion in \citealt{Kruijssen2014,Kruijssen2015}).

\subsection{Polarization dilution by the GC diffuse plasma emission and detectability with modern instruments}
\label{Main:IXPE}

\begin{table*}
\begin{center}
   \begin{tabular}{ccccccc}
   \hline
      Molecular cloud	& P (\%) 	& $\psi$ ($^\circ$)	& $f_\mathrm{R}$ (\%)     & 
      P$_\mathrm{exp.}$ (\%) 	& P$_{\rm detect.}$ (\%) 	& $\psi_{\rm detect.}$ ($^\circ$)\\
   \hline
      Sgr~B2    	& 65.0			& 88.3			& 70.0		& 45.5		& 57.4$~\pm~$4.4		& 83.3$~\pm~$3.4	\\      
      Sgr~B1    	& 76.9			& 84.4			& 52.6		& 40.5		& 40.4$~\pm~$3.9		& 80.3$~\pm~$3.3	\\
      G0.11-0.11    	& 55.8			& 61.6			& --		& --		& --				& --			\\
      Bridge~E    	& 12.7			& 67.9			& --		& --		& --				& --			\\
      Bridge~D    	& 0.1			& 74.2			& --		& --		& --				& --			\\
      Bridge~B2    	& 15.8			& 77.8			& --		& --		& --				& --			\\
      MC2    		& 25.8			& 73.8			& --		& --		& --				& --			\\     
      MC1    		& 0.1			& 77.5			& --		& --		& --				& --			\\
      Sgr~C3    	& 32.9			& 106.4			& 50.7		& 16.7		& 15.5$~\pm~$2.4		& 109.0$~\pm~$4.5	\\    
      Sgr~C2    	& 34.9			& 99.1			& 63.0		& 22.0		& 17.9$~\pm~$3.8		& 99.1$~\pm~$5.6	\\
      Sgr~C1    	& 31.1			& 94.6			& 60.2		& 18.7		& 23.1$~\pm~$3.3		& 98.1$~\pm~$6.0	\\
   \hline
   \end{tabular}
   \caption{Integrated 4 -- 8~keV polarization degree $P$ of the reflection component (including neutral iron lines) and 
	   polarization position angle $\psi$ of the GC molecular clouds from the simulation with {\sc stokes}. Polarization 
	   angles are defined with respect to Galactic North, with positive defined as West to North. The fraction of the 
	   total flux that is reflected $f_\mathrm{R}$ is computed from \citet{Ryu2009} and \citet{Ryu2013}, allowing us to 
	   evaluate the diluted polarization signal P$_\mathrm{exp.}$. Using Monte Carlo simulations associated with the GPD
	   instrument (see text), we finally show estimations of the polarization degree P$_{\rm detect.}$ and angle 
	   $\psi_{\rm detect.}$ that a future polarimeter would detect. The empty cells correspond to clouds with too 
	   low X-ray luminosities to be observed within 3 Ms or with unestimated fractions of the reflected flux.}
\end{center}
  \label{Tab:MDP_est}
\end{table*}

\begin{figure*}
  \centering
  \includegraphics[trim = 20mm 50mm 15mm 50mm, clip, width=19cm]{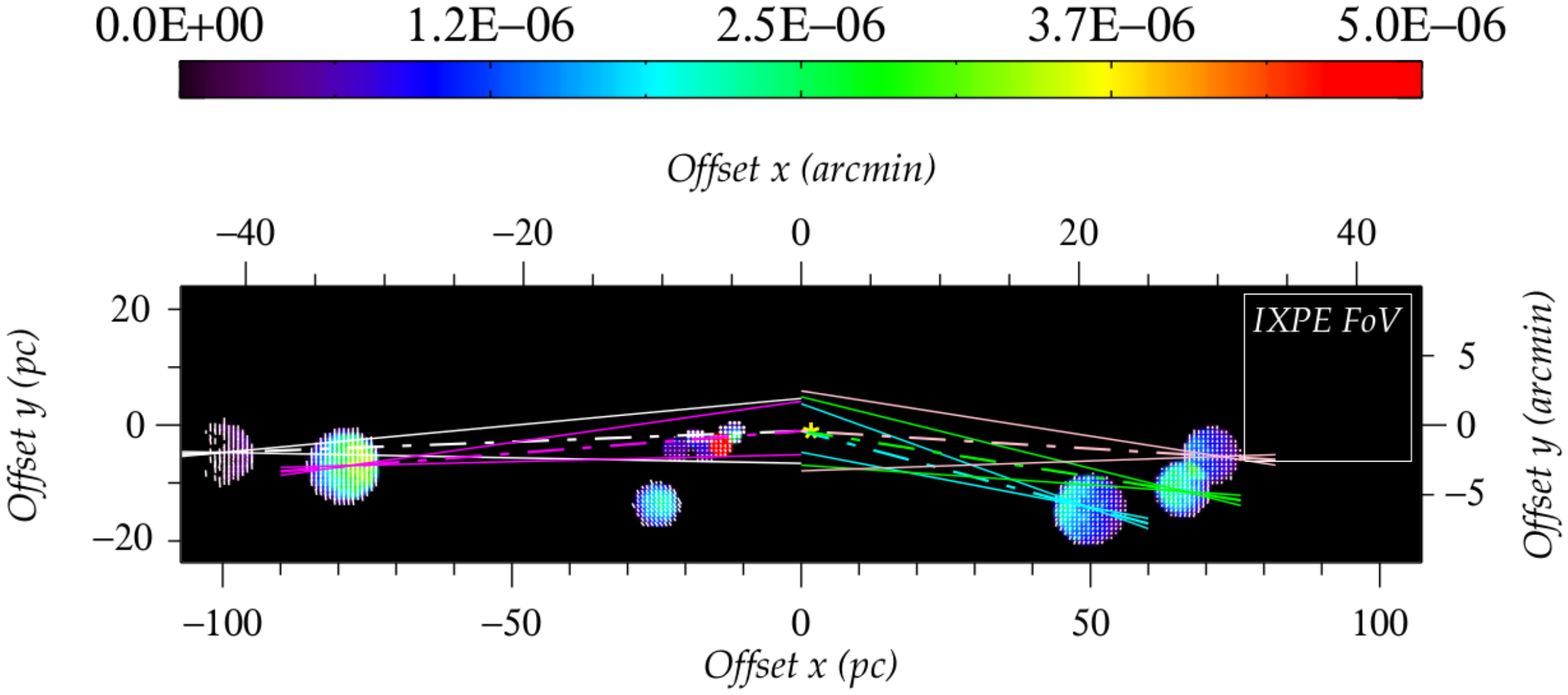} 
  \caption{Integrated polarization image of the GC showing how the angle of polarization 
	   would constrain the position of the illuminating source. The polarized flux 
	   is shown, color-coded and with arbitrary units. The field of view (FoV) of 
	   our test case {\it IXPE} is indicated with a white box, and a yellow star 
	   indicates the position of Sgr~A$^*$. Colored segments (Sgr~B2: white; Sgr~B1: magenta; 
	   Sgr~C3: cyan; Sgr~C2: green; Sgr~C1: pink) are representative of the 
	   estimated polarization position angle (dashed line) and its associated 
	   error (solid line).} 
  \label{Fig:Segments} 
\end{figure*}

To evaluate how modern imaging polarimeters may constrain the angular position of the source which illuminated the GC molecular clouds in 
the past, we simulate their observations taking into account the complex environment in which these sources are immersed. One of the most elaborate, 
technologically-ready X-ray polarimeter is the Gas Pixel Detector (GPD, \citealt{Costa2001,Bellazzini2006,Bellazzini2010}). The GPD is particularly 
sensitive to the X-ray polarization in the 2 -- 10 keV energy range, also offering fine location accuracy and moderate energy resolution 
\citep{Muleri2010,Fabiani2014}. These characteristics are very well matched with the required moderate angular resolution of 4 -- 5~arcmin for 
performing these observations (see Fig.~\ref{Fig:Sketch}).

As a test case of the GPD in the context of modern polarimetric missions, we rely on the imaging capabilities of {\it IXPE} (the Imaging X-ray 
Polarimetry Explorer, a mission concept to be proposed to the next NASA/SMEX call). The {\it IXPE}'s 30~arcsec half-power diameter
roughly corresponds to the spatial resolution of our images, so that a comparison between the resolution of the instrument and our simulation 
is straightforward. Equipped with GPDs, such a mission will allow to single 
out and remove the contribution of any point-like sources, even if transient (namely, Sgr A* flares and transients), which may be active during 
the observation; therefore we can safely neglect any contamination from those sources. Nonetheless, we have to account for the diffuse Galactic 
plasma emission that is expected to be unpolarized and in any case not correlated with the position of the illuminating source. Therefore, the 
plasma contribution has to be subtracted from the flux coming from the molecular cloud; alternatively, the simulated polarization has to be diluted, 
with respect to the values presented in the previous section, by an amount which depends on what fraction of the total flux is due to the reflected 
component. While during flight we could always compare the results from these two different methods, in this paper we choose the latter for 
practical reasons.

We estimate the plasma and the reflected contributions for the molecular clouds in the Sgr~B and Sgr~C complexes by means of the spectral 
decomposition performed by \citet{Ryu2009} and \citet{Ryu2013}, respectively. In these works, the spectra of Sgr~B1, Sgr~B2, 
Sgr~C1, Sgr~C2 and Sgr~C3 are each fitted with two spectral components which separately take into account the plasma contribution 
and the emission due to the reflection of the external source radiation. Labeling these two spectral components $F_\mathrm{plasma}(E)$ 
and $F_\mathrm{refl.}(E)$, the ``dilution'' factor $f_\mathrm{R}$ by which we multiply the polarization presented above to obtain the 
expected degree polarization $P_\mathrm{exp.}$ is:
\begin{equation}
f_\mathrm{R} = \frac{\int_{4~\mathrm{keV}}^{8~\mathrm{keV}} F_\mathrm{refl.}(E) \mathrm{d}E}{\int_{4~\mathrm{keV}}^{8~\mathrm{keV}} 
\left[ F_\mathrm{refl.}(E) + F_\mathrm{plasma}(E)\right]\mathrm{d}E},
\end{equation}
where $E$ is the energy. The energy interval 4 -- 8~keV is chosen to maximize the reflected contribution in the energy range where IXPE is 
most sensitive. The dilution $f_\mathrm{R}$ for the different clouds is in the range 50\% -- 70\% assuming the best fit parameters 
estimated by \citet{Ryu2009} and \citet{Ryu2013} (see Tab.~2); however, the uncertainty on this value depends on the uncertainties on 
the parameters of the fit deconvolution. For example, changing such parameters in the 90\% confidence level in case of Sgr~B2 results in a 
value of $f_\mathrm{R}$ between about 64\% and 75\%, with a mean value of 70\% which coincides with the number reported in Tab.~2. 
Therefore, there is a systematic uncertainty on the expected polarization of the order of about 10\% of its value, but this does not affect 
our ultimate goal, which is to explore the feasibility of the polarization measurement with reasonable assumptions. Moreover, it is 
reasonable to assume that when the measurement will be actually done, such a systematic uncertainty will be reduced by the more accurate 
measurements carried out by other spectroscopy-dedicated satellites.

The inputs for the Monte Carlo routine (described in detail in \citealt{Dovciak2011}), are the net polarization $P_\mathrm{exp.}$, 
reported in Tab.~2, and the flux of each molecular cloud. This returns an estimate of the measured polarization for the selected observation 
time\footnote{Here we do not need to consider effects of general relativity on polarization of light near a black hole, because the assumed 
scattering clouds are located relatively far from the event horizon.}. The GPD field of view is sufficient to observe the Sgr~B and Sgr~C 
complexes in a single pointing each; therefore, we assumed to carry out a single 1~Ms long observation for Sgr~B and another single 2~Ms 
long observation of Sgr~C, whose expected degree of polarization is lower because of the less favorable scattering geometry. We also 
estimated the detector's residual background rate, based on past flown gas detectors with similar gas mixture. Thanks to the GPD's 
imaging capabilities, this is about 10 times smaller than the expected signal from the reflected component of the molecular clouds. The results 
are reported in Tab.~2 and shown graphically in Fig.~\ref{Fig:Segments}. In this picture, we reported the polarization angles $\psi_{\rm 
detect.}$ such as they would be measured by a modern imaging polarimeter with a $1\sigma$ error of a few degrees for each cloud. 
Fig.~\ref{Fig:Segments} demonstrates that the measurement of $\psi$ would allow us to constrain the angular position of the illuminating source 
very tightly. Moreover, the five molecular clouds provide as many independent constraints, so it is clear that a future X-ray polarimetric 
satellite equipped with a mapping instrument would be able to test unambiguously the scattering origin of the X-ray emission from GC 
molecular clouds. In principle, all the other mentioned molecular clouds could be observed (e.g. G0.11-0.11, the Bridge, MC1 or MC2) with a 
single pointing. However, due to their lower expected net polarization, as well as the increased contribution of the plasma emission, the 
observation strategy for those reflection nebulae will be driven by the results obtained for the Sgr~B and the Sgr~C complexes.

\section{Concluding remarks}
\label{Conclusion}

To probe the crowded field of the GC, an instrument with imaging capability is essential. X-ray polarimetry is 
needed to test (in a novel way) the physical processes operating near the Galactic supermassive black hole.

In this paper, we simulated the 4 to 8~keV polarization response of the observed, 6.4~keV bright giant 
molecular regions in the GC to a Sgr~A$^*$ flaring event. We found that the scattering nebulae present 
a variety of polarization signatures, ranging from nearly unpolarized to highly polarized (with $P \sim$ 77\%) 
fluxes. The brightness distribution of the reprocessed flux compared with the contours of the spherical clumps is in 
agreement with past observations and tends to point towards a flaring scenario to explain the detection of hard 
X-ray spectra and prominent iron K$\alpha$ fluorescence features. Future observations would be able to test our 
predictions against an alternative mechanism proposed to explain the same X-ray signatures by low-energy cosmic-ray 
electron interactions with neutral matter \citep{Valinia2000,Yusef2002}. This scenario, not specifically excluded by 
observations \citep{Capelli2011}, suggests that the resulting X-ray power-law originates from thermal bremsstrahlung 
emission, and thus the net polarization would be either null for an isotropic distribution of electrons, or at least 
different from Compton scattering-induced polarization. The key feature needed to discriminate between the two scenarios
is to measure the angle of polarization. Indeed, in comparison with the observed degree of polarization affected by 
the Galactic plasma and by its characteristics, the polarization position angle of a photon will not suffer any GC 
plasma-induced rotation along its journey towards Earth.

To assess the validity of the flaring hypothesis, we simulated an observation of the reflection nebulae with the GPD, 
a modern imaging polarimeter to be mounted on future X-ray polarimetric satellites, taking into account the presence of a 
diffuse, unpolarized, plasma emission towards the GC. While such an effect decreases the amount of polarization, 
we found that with a 1~Ms observation of the Sgr~B complex and/or with a 2~Ms observation of the Sgr~C complex, 
the polarization imager of a future instrument would be able to unambiguously determine the history of Sgr~A$^*$ by 
pinpointing the source of the primary emission.

In this context, the presence of 7 transient X-ray binaries within 23~pc of the GC (4 within 1~pc, \citealt{Muno2005}) 
could be a challenge for future observations since a past X-ray outburst of one of these sources could have mimicked
a Sgr~A$^*$ flare. However, since those objects are likely low-mass X-ray binaries (\textit{ibid.}), their putative 
past outburst would hardly exceed 10$^{37 - 38}$~ergs.s$^{-1}$ (assuming L$_X$ = 0.1$\dot{M}_{\rm peak}$c$^2$,
\citealt{Dubus2001}), which is still two orders of magnitude lower than the expected light echo of Sgr~A$^*$
($L_X >$~10$^{39}$~erg.s$^{-1}$). In addition, Fig.\ref{Fig:Segments} shows that modern imaging polarimeters are able 
to constrain the emitting source to within less than 10~pc around the SMBH, removing half of the X-ray transients.

\acknowledgements The authors would like to thank the anonymous referee for useful and constructive comments.
This research has been partially supported by the European Union Seventh Framework Programme (FP7/2013–2017)
under grant agreement no.~312789, StrongGravity. FM thanks the grant COST-CZ LD12010 for additional funding.
VK and DK are grateful to the Czech Science Foundation - Deutsche Forschungsgemeinschaft collaboration project 
(GACR 13-00070J).



\begin{thebibliography}{}
\bibitem[\protect\citeauthoryear{An et al.}{2013}]{An2013} An, D., Ram{\'{\i}}rez, S.~V., \& Sellgren, K.\ 2013, ApJs, 206, 20 
\bibitem[\protect\citeauthoryear{Baganoff et al.}{2001}]{Baganoff2001} Baganoff, F.~K., Bautz, M.~W., Brandt, W.~N., et al.\ 2001, Nature, 413, 45 
\bibitem[\protect\citeauthoryear{Bellazzini et al.}{2006}]{Bellazzini2006} Bellazzini, R., Angelini, F., Baldini, L., et al.\ 2006, Nuclear Instruments and Methods in Physics Research A, 560, 425
\bibitem[\protect\citeauthoryear{Bellazzini \& Muleri}{2010}]{Bellazzini2010} Bellazzini, R., \& Muleri, F. 2010, Nuclear Instruments and Methods in Physics Research A, 623, 766 
\bibitem[\protect\citeauthoryear{Capelli et al.}{2011}]{Capelli2011} Capelli, R., Warwick, R.~S., Porquet, D., Gillessen, S., \& Predehl, P.\ 2011, A\&A, 530, AA38 
\bibitem[\protect\citeauthoryear{Capelli et al.}{2012}]{Capelli2012} Capelli, R., Warwick, R.~S., Porquet, D., Gillessen, S., \& Predehl, P.\ 2012, A\&A, 545, A35 
\bibitem[\protect\citeauthoryear{Churazov et al.}{2002}]{Churazov2002} Churazov, E., Sunyaev, R., \& Sazonov, S.\ 2002, MNRAS, 330, 81
\bibitem[\protect\citeauthoryear{Clavel et al.}{2013}]{Clavel2013} Clavel, M., Terrier, R., Goldwurm, A., et al.\ 2013, A\&A, 558, A32 
\bibitem[\protect\citeauthoryear{Costa et al.}{2001}]{Costa2001} Costa, E., Soffitta, P., Bellazzini, R., et al.\ 2001, Nature, 411, 662 
\bibitem[\protect\citeauthoryear{Dov{\v c}iak et al.}{2011}]{Dovciak2011} {Dov{\v c}iak}, M. and {Muleri}, F. and {Goosmann}, R.~W., et al.\ 2011, ApJ, 731, 75
\bibitem[\protect\citeauthoryear{Downes et al.}{1980}]{Downes1980} Downes, D., Wilson, T.~L., Bieging, J., \& Wink, J.\ 1980, A\&As, 40, 379 
\bibitem[\protect\citeauthoryear{Dubus et al.}{2001}]{Dubus2001} Dubus, G., Hameury, J.-M., \& Lasota, J.-P.\ 2001, A\&A, 373, 251 
\bibitem[\protect\citeauthoryear{Fabiani et al.}{2014}]{Fabiani2014} Fabiani, S., Costa, E., Del Monte, E., et al.\ 2014, ApJS, 212, 25
\bibitem[\protect\citeauthoryear{Ghez et al.}{2008}]{Ghez2008} Ghez, A.~M., Salim, S., Weinberg, N.~N., et al.\ 2008, ApJ, 689, 1044 
\bibitem[\protect\citeauthoryear{Goosmann \& Gaskell}{2007}]{Goosmann2007} Goosmann, R.~W., \& Gaskell, C.~M.\ 2007, A\&A, 465, 129
\bibitem[\protect\citeauthoryear{Inui et al.}{2009}]{Inui2009} Inui, T., Koyama, K., Matsumoto, H., \& Tsuru, T.~G.\ 2009, PASJ, 61, 241 
\bibitem[\protect\citeauthoryear{Koyama et al.}{1986}]{Koyama1986} Koyama, K., Makishima, K., Tanaka, Y., \& Tsunemi, H.\ 1986, PASJ, 38, 121 
\bibitem[\protect\citeauthoryear{Koyama et al.}{1989}]{Koyama1989} Koyama, K., Awaki, H., Kunieda, H., Takano, S., \& Tawara, Y.\ 1989, Nature, 339, 603 
\bibitem[\protect\citeauthoryear{Koyama et al.}{1996}]{Koyama1996} Koyama, K., Maeda, Y., Sonobe, T., et al.\ 1996, PASJ, 48, 249 
\bibitem[\protect\citeauthoryear{Koyama et al.}{2007}]{Koyama2007} Koyama, K., Hyodo, Y., \& Inui, T., et al.\ 2007, PASJ, 59, 245
\bibitem[\protect\citeauthoryear{Kruijssen et al.}{2014}]{Kruijssen2014}  Kruijssen, J.~M.~D., Longmore, S.~N., Elmegreen, B.~G., et al.\ 2014, MNRAS, 440, 3370
\bibitem[\protect\citeauthoryear{Kruijssen et al.}{2015}]{Kruijssen2015} Kruijssen, J.~M.~D., Dale, J.~E., \& Longmore, S.~N.\ 2015, MNRAS, 447, 1059 
\bibitem[\protect\citeauthoryear{Marin et al.}{2012}]{Marin2012} Marin, F., Goosmann, R.~W., Gaskell, C.~M., Porquet, D., \& Dov{\v c}iak, M.\ 2012, A\&A, 548, A121 
\bibitem[\protect\citeauthoryear{Marin et al.}{2014}]{Marin2014} Marin, F., Karas, V., Kunneriath, D., \& Muleri, F.\ 2014, MNRAS, 441, 3170
\bibitem[\protect\citeauthoryear{Marin \& Dov{\v c}iak}{2015}]{Marin2015} Marin, F., \& Dov{\v c}iak, M.\ 2015, A\&A, 573, AA60 
\bibitem[\protect\citeauthoryear{Mewe}{1999}]{Mewe1999} Mewe, R.\ 1999, X-Ray Spectroscopy in Astrophysics, 520, 109
\bibitem[\protect\citeauthoryear{Meyer et al.}{2011}]{Meyer2011} Meyer-Hofmeister, E., \& Meyer, F.\ 2011, A\&A, 527, A127 
\bibitem[\protect\citeauthoryear{Molinari et al.}{2011}]{Molinari2011} Molinari, S., Bally, J., Noriega-Crespo, A., et al.\ 2011, ApJL, 735, L33
\bibitem[\protect\citeauthoryear{Muleri et al.}{2010}]{Muleri2010} Muleri, F., Soffitta, P., Baldini, L., et al.\ 2010, Nuclear Instruments and Methods in Physics Research A, 620, 285
\bibitem[\protect\citeauthoryear{Muno et al.}{2005}]{Muno2005} Muno, M.~P., Pfahl, E., Baganoff, F.~K., et al.\ 2005, ApJl, 622, L113
\bibitem[\protect\citeauthoryear{Murakami et al.}{2001a}]{Murakami2001a} Murakami, H., Koyama, K., \& Maeda, Y.\ 2001a, ApJ, 558, 687 
\bibitem[\protect\citeauthoryear{Murakami et al.}{2001b}]{Murakami2001b} Murakami, H., Koyama, K., Tsujimoto, M., Maeda, Y., \& Sakano, M.\ 2001b, APJ, 550, 297 
\bibitem[\protect\citeauthoryear{Nowak et al.}{2012}]{Nowak2012} Nowak, M.~A., Neilsen, J., Markoff, S.~B., et al.\ 2012, ApJ, 759, 95 
\bibitem[\protect\citeauthoryear{Ponti et al.}{2010}]{Ponti2010} Ponti, G., Terrier, R., Goldwurm, A., Belanger, G., \& Trap, G.\ 2010, ApJ, 714, 732
\bibitem[\protect\citeauthoryear{Ponti et al.}{2014}]{Ponti2014} Ponti, G., Morris, M.~R., Clavel, M., et al.\ 2014, IAU Symposium, 303, 333
\bibitem[\protect\citeauthoryear{Porquet et al.}{2003}]{Porquet2003} Porquet, D., Predehl, P., Aschenbach, B., et al.\ 2003, A\&A, 407, L17 
\bibitem[\protect\citeauthoryear{Porquet et al.}{2008}]{Porquet2008} Porquet, D., Grosso, N., Predehl, P., et al.\ 2008, A\&A, 488, 549 
\bibitem[\protect\citeauthoryear{Revnivtsev et al.}{2009}]{Revnivtsev2009} Revnivtsev, M., Sazonov, S., Churazov, E., et al.\ 2009, Nature, 458, 1142
\bibitem[\protect\citeauthoryear{Ryu et al.}{2009}]{Ryu2009} Ryu, S.~G., Koyama, K., Nobukawa, M., Fukuoka, R., \& Tsuru, T.~G.\ 2009, PASJ, 61, 751 
\bibitem[\protect\citeauthoryear{Ryu et al.}{2013}]{Ryu2013} Ryu, S.~G., Nobukawa, M., Nakashima, S., et al.\ 2013, PASJ, 65, 33
\bibitem[\protect\citeauthoryear{Shi et al.}{2006}]{Shi2006} Shi, Y., Rieke, G.~H., Hines, D.~C., et al.\ 2006, ApJ, 653, 127 
\bibitem[\protect\citeauthoryear{Sidoli \& Mereghetti}{1999}]{Sidoli1999} Sidoli, L., \& Mereghetti, S.\ 1999, A\&A, 349, L49
\bibitem[\protect\citeauthoryear{Sunyaev et al.}{1993}]{Sunyaev1993} Sunyaev, R.~A., Markevitch, M., \& Pavlinsky, M.\ 1993, ApJ, 407, 606 
\bibitem[\protect\citeauthoryear{Sunyaev \& Churazov}{1998}]{Sunyaev1998} Sunyaev, R., \& Churazov, E.\ 1998, MNRAS, 297, 1279 
\bibitem[\protect\citeauthoryear{Valinia et al.}{2000}]{Valinia2000} Valinia, A., Tatischeff, V., Arnaud, K., Ebisawa, K., \& Ramaty, R.\ 2000, ApJ, 543, 733 
\bibitem[\protect\citeauthoryear{Weisskopf et al.}{2008}]{Weisskopf2008} Weisskopf, M.~C., Bellazzini, R., Costa, E., et al.\ 2008, Proceedings of the SPIE, 7011
\bibitem[\protect\citeauthoryear{Weisskopf et al.}{2014}]{Weisskopf2014} Weisskopf, M.~C., Bellazzini, R., Costa, E., et al.\ 2014, AAS/High Energy Astrophysics Division, 14, \#116.15  
\bibitem[\protect\citeauthoryear{Yusef-Zadeh et al.}{2002}]{Yusef2002} Yusef-Zadeh, F., Law, C., \& Wardle, M.\ 2002, ApJL, 568, L121 
\end{thebibliography}
\end{document}